\documentclass[12pt,preprint]{aastex}
\newcommand{\etal}{{\it et al.}}

\newcommand{\gsim}{\ga}

\newcommand{\ra}[3]{#1$^{\rm h}$#2$^{\rm m}$#3$^{\rm s}$}
\newcommand{\dec}[3]{$#1^\circ#2{^\prime}#3^{\prime\prime}$}
\newcommand{\grb}{GRB\thinspace{000210}}

\newcommand{\ha}{\rm cm^{-2}}

\newcommand{\Msun}{M_\sun}

\newcommand{\rcm}{\rm cm}

\newcommand{\flux}{\rm erg\ cm^{-2}\ s^{-1}}

\newcommand{\csq}{\chi^2}

\newcommand{\ltsim}{\la}
\begin{document}

\title{The bright  Gamma-Ray Burst of February 10, 2000: a case study
of an optically dark GRB}

% Preliminary author list
\author{
L. Piro\altaffilmark{1},
 D. A. Frail\altaffilmark{2},
 J.Gorosabel\altaffilmark{3,4},
 G. Garmire\altaffilmark{5},
 P. Soffitta\altaffilmark{1},
 L. Amati\altaffilmark{6},
 M.I. Andersen\altaffilmark{7},
 L. A. Antonelli\altaffilmark{8}
 E.Berger\altaffilmark{9},
 F. Frontera\altaffilmark{6,10},
 J. Fynbo\altaffilmark{11},
 G. Gandolfi\altaffilmark{1},
 M.R. Garcia\altaffilmark{12},
 J. Hjorth\altaffilmark{13},
 J. in 't Zand\altaffilmark{14}
 B.L. Jensen\altaffilmark{13},
 N.Masetti\altaffilmark{6},
 P. M\o ller\altaffilmark{11},
 H. Pedersen\altaffilmark{13},
 E. Pian\altaffilmark{15},
 M. H. Wieringa\altaffilmark{16}
}


\begin{abstract}

  The gamma-ray burst \grb\ had the highest $\gamma$-ray peak flux of
  any  event localized by {\em BeppoSAX} as yet but it did not have a
  detected optical afterglow, despite prompt and deep searches down to
  $R_{lim}\approx 23.5$. It is therefore one of the events recently
  classified as dark GRBs, whose origin is still unclear.  {\em Chandra}
  observations allowed us to localize the X-ray afterglow of \grb\ to
  within $\approx 1$\arcsec\ and a radio transient was detected with
  the VLA. The precise X-ray and radio positions allowed us to
  identify the likely host galaxy of this burst, and to measure its
  redshift, $z=0.846$. The probability that this galaxy is a field
  object is $\approx 1.6\times 10^{-2}$. The X-ray spectrum of the afterglow
  shows significant absorption in excess of the Galactic one
  corresponding, at the redshift of the
  galaxy, to $N_H=(5\pm1)\times 10^{21}$ cm$^{-2}$.  The amount of dust
  needed to absorb the optical flux of this object is consistent with
  the above HI column density, given a dust-to-gas ratio
  similar to that of our Galaxy.  We do not find
  evidence for a partially ionized absorber expected if the absorption
  takes place in a Giant Molecular Cloud. We therefore conclude that
  either the gas is local to the GRB, but is condensed in small-scale
  high-density ($n\gtrsim10^9$ cm$^{-3}$) clouds, or that the GRB is
  located in a dusty, gas-rich region of the galaxy.  Finally, we
  examine the hypothesis that \grb\ lies at $z\gtrsim5$ (and
  therefore that the optical flux is extinguished by Ly$_\alpha$ forest clouds),
   but we conclude that the
  X-ray absorbing medium would have to be substantially thicker from
  that observed in GRBs with optical afterglows.

 % We do
 % not find evidence of ionization of the X-ray absorbing gas, that
 % would be expected if the GRB exploded in a giant molecular
 % cloud (GMC). The X-ray  absorption can still be consistent with this scenario,
 % if produced by  small-scale high-density ($n\gtrsim10^8$cm$^{-3}$) fluctuations
 %  of the medium in the GMC.
 % This information, together with the properties of the galaxy
 % and the location of the GRB, suggest that the absorption
 % could also originate in the intergalactic medium of the host galaxy.

\end{abstract}

\altaffiltext{1}{Istituto Astrofisica Spaziale \& Fisica Cosmica,
C.N.R., Via Fosso del Cavaliere, 00133 Roma, Italy}

\altaffiltext{2}{National Radio Astronomy Observatory, Socorro,
NM, 87801, USA}

\altaffiltext{3}{ Danish Space Research Institute, Juliane Maries
Vej 30, DK--2100 Copenhagen \O, Denmark}
%           jgu@dsri.dk

\altaffiltext{4}{Instituto de Astrof\'{\i}sica de Andaluc\'{\i}a,
CSIC, Apdo. Correos 3004, 18080 Granada, Spain}

\altaffiltext{5}{Department of Astronomy and Astrophysics, 525
Davey Lab, Penn State University, University Park, PA 16802, USA}

\altaffiltext{6}{Istituto Astrofisica Spaziale \& Fisica Cosmica,
sezione Bologna, C.N.R.,Via Gobetti 101, 40129 Bologna, Italy}

\altaffiltext{7}{Division of Astronomy, University of Oulu P.O.
Box 3000, FIN-90014 University of Oulu, Finland}
%            michael.andersen@oulu.fi

\altaffiltext{8}{Osservatorio Astronomico Roma, INAF, Via Frascati
33, 00040  Monte Porzio Catone, Roma, Italy}

\altaffiltext{9}{California Institute of Technology, Palomar
  Observatory 105-24, Pasadena, CA 91125}

\altaffiltext{10}{Dip. Fisica, Universita' Ferrara, Via Paradiso
12, Ferrara, Italy}

\altaffiltext{11}{European Southern Observatory,
Karl--Schwarzschild--Stra\ss e 2, D--85748 Garching, Germany}
%           jfynbo@eso.org,pmoller@eso.org

\altaffiltext{12}{Harvard-Smithsonian Center for Astrophysics, 60
Garden St. Cambridge, MA 02138, USA}

\altaffiltext{13}{Astronomical Observatory, University of
Copenhagen, Juliane Maries Vej 30, DK--2100 Copenhagen \O,
Denmark}
%           jens@astro.ku.dk,holger@astro.ku.dk,brian\_j@astro.ku.dk

\altaffiltext{14}{ Space Research Organization in the Netherlands,
Sorbonnelaan 2, 3584 CA Utrecht, The Netherlands}

\altaffiltext{15}{Osservatorio Astronomico Trieste, INAF, Via G.
Tiepolo 11, I-34131 Trieste, Italy}

\altaffiltext{16}{Paul Wild Observatory, Locked Bag 194, Narribri
NSW  2390, Australia}

%\altaffiltext{5}{Columbia University, Dept. of Astrophysics, 545 W. 114th St., New York, NY 10027, USA}

%\altaffiltext{7}{UC Space Science Laboratory, Berkeley, CA 94720-7450, USA}

\newpage
\section{Introduction}\label{sec:intro}

It is observationally well-established that about half of
accurately localized  gamma-ray bursts (GRBs) do not produce a
detectable optical afterglow \citep{fkw+00,fjg+01}, while most of
them ($\approx90\%$) have an X-ray afterglow \citep{piro01}.
Statistical studies have shown that the optical searches of these
events, known variously as ``dark GRBs'', ``failed optical
afterglows'' (FOA), or ``gamma-ray bursts hiding an optical
source-transient'' (GHOST), have been carried out to
  magnitude limits fainter on average than the  known sample of
optical afterglows (\citet{lcg01,ry01}, but see also
\citet{fjg+01}). Some of these GRBs could be intrinsically faint
events, but this fraction cannot be very high, because the
majority of dark GRBs shows the presence of an X-ray afterglow
similar to that observed in GRBs with optical afterglows
\citep{piro01,lcg01}. Thus dark bursts could constitute a distinct
class of events and not only be the result of an inadequate
optical search, but it is unclear whether this observational
property derives from a  single origin or it is a combination of
different causes.

If the progenitors of long-duration GRBs are massive stars
\citep{pac98b}, as current evidence suggests ({\em e.g.,}
\citet{bkd+99,pgg+00}), extinction of optical flux by dusty
star-forming regions is likely to occur for a substantial fraction
of events ({\it the obscuration scenario}). Another possibility is
that dark GRBs are located at redshift $z\gsim$5, with the optical
flux being absorbed by the intervening Ly$\alpha$ forest clouds
({\it the
  high-redshift scenario}).

Dark bursts which can be localized to arcsecond accuracy, through
a detection of either   their X-ray or radio afterglow, are of
particular interest.  The first and best-studied example was
GRB\,970828 for which prompt, deep searches down to R$\sim$24.5
failed to detect an optical afterglow \citep{odk+97, ggv+98d}
despite it was localized within a region of only 10\arcsec\ radius
by the ROSAT satellite \citep{gse+97}. \citet{dfk+01} recently
showed how the detection of a short-lived radio transient for
GRB\, 970828 allowed them to identify the probable host galaxy and
to infer its properties (redshift, luminosity and morphology). In
addition, they used estimates of the column density of absorbing
gas from X-ray data, and lower limits on the rest frame extinction
(A$_V>3.8$) to quantify the amount of obscuration towards the GRB.

Given the extreme luminosity of GRBs and their probable
association with massive stars, it is expected that some fraction
of events will be located beyond $z>5$ \citep{lr00}. These would
be probably classified as dark bursts because the UV light, which
is strongly attenuated by absorption in the Ly$_\alpha$ forest, is
redshifted into the optical band. Fruchter (1999)\nocite{fru99}
first suggested such an explanation for the extreme red color of
the optical/NIR emission for GRB\,980329, although an alternative
explanation based on   H$_2$ absorption in the GRB environment
would imply a somewhat lower redshift \citep{draine00}. In a
recent paper \citet{jah+02} derive a photometric redshift
$z\approx3.5$. We note that the three redshifts determined or
suggested so far for dark GRBs ($z=0.96$, GRB970828,
\citet{dfk+01};$z=1.3$, GRB990506, \citet{tbf+00, bkd02};
$z\approx0.47$, GRB000214, \citet{apv+00}) are in the range of
those measured for most bright optical afterglows, but whether
this applies to the majority of these events is still to be
assessed. Particularly interesting in this respect is the case of
the  so-called X-ray flashes or  X-ray rich GRBs discovered by
{\em BeppoSAX}\citep{hzkw01}. In most of these events no optical
counterpart has been found. The only tentative association claimed
as yet is for the event of Oct.30, 2001, where a candidate host
galaxy of magnitude V$\approx25$ has been found
 in the direction of the afterglow \citep{fpk+02}. We note,
however, that  the probability that this object is a foreground
galaxy is not negligible ($P\approx3\times10^{-2}$, see e.g.
eq.\ref{eq:p2} in Sect.2.3).
 The {\it high-redshift scenario} would naturally explain both
the absence of optical counterpart and the high-energy spectrum,
because the peak of the gamma-ray spectrum would be redshifted
into the X-ray band.

%{\bf something more here}

If we are to use dark bursts to study obscured star formation in
the universe \citep{dfk+01}, we must first understand the source
of the extinction. For those afterglows which are not at $z>5$  it
is important to establish whether they are dark because of a dense
circumburst medium \citep{ry01}, or  their optical emission is
extinguished by line-of-sight absorption from the medium of the
host galaxy. We can use the properties of the afterglow, its
location within the host galaxy, and the properties of the host
galaxy itself to address this question. In this paper, we report
observations of the burst \grb\ which was discovered and localized
by {\em BeppoSAX}.  A {\em Chandra} observation of the {\em
  BeppoSAX} error box enabled us to localize the likely host galaxy
of the event and to identify a short-lived radio-transient,
further refining the position to sub-arcsec accuracy.  From
sensitive upper limits on the absence of an optical afterglow we
estimate the amount of extinction by dust and from the X-ray
spectrum the amount of absorbing gas. \grb\ appears to be the
newest member of a small but growing group of well-localized dark
bursts \citep{fkb+99,tbf+00,dfk+01}.

\section{Observations}\label{sec:obs}

\subsection{Gamma-Ray and X-ray Observations}\label{sec:grb}

\begin{figure}
\epsscale{0.7}
%\plotone{grbpap7.ps}
\plotone{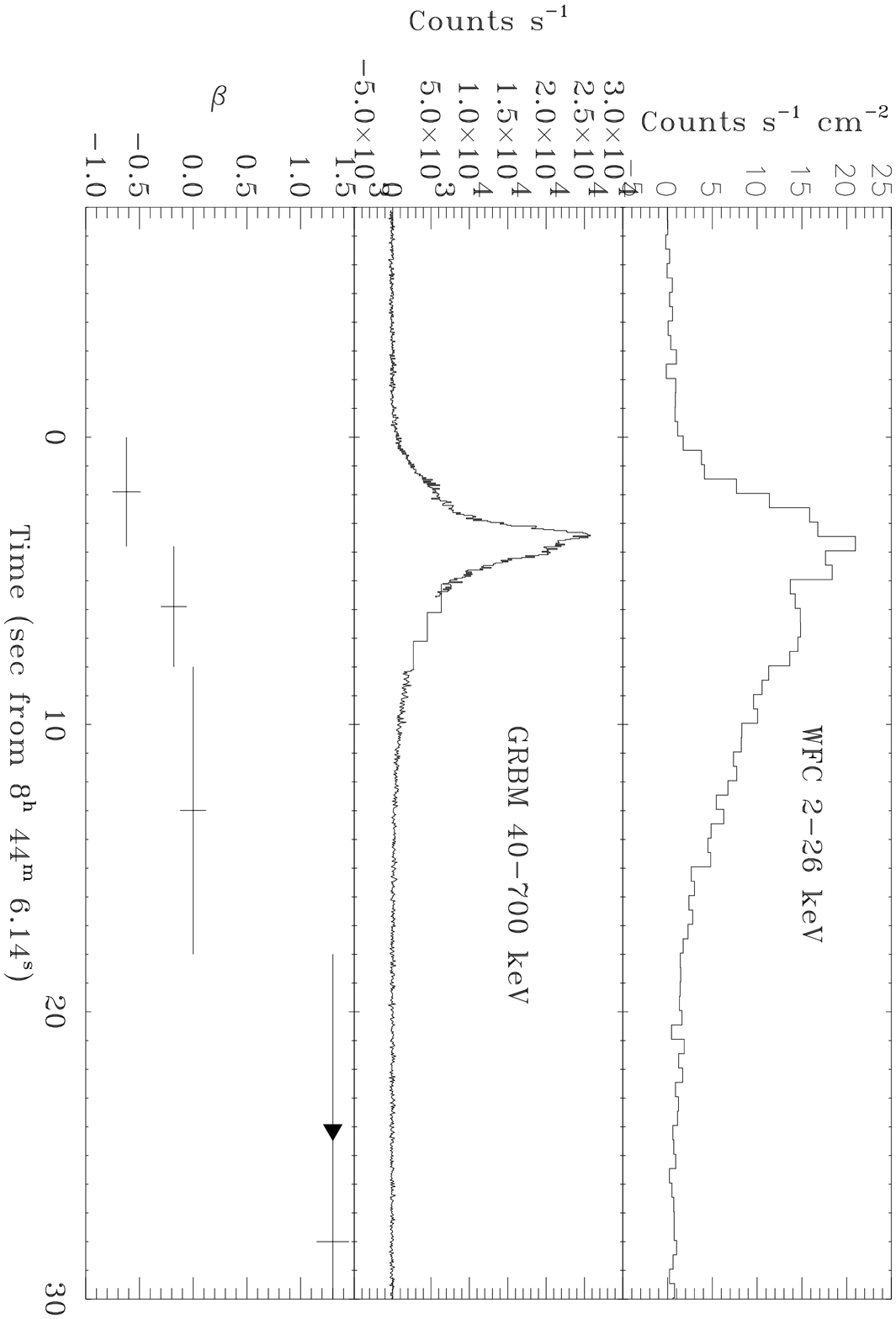}
%\plotone{test.ps}
\caption{Light curves of GRB\,000210 in the {\em BeppoSAX} WFC (upper
  panel), GRBM (middle panel) and energy spectral index
  ($\beta=\Gamma-1$, $F\propto E^{-\beta}$) evolution in the
  WFC (lower panel). The gap between 5.5 and 8 sec in the
  GRBM data is due to a telemetry loss. In that interval we have
  plotted the 1 sec resolution data of the ratemeter of the
  instrument, while the other data points have a time bin of 31.25
  ms.
  The last point of the spectral index is relative to the interval 18-80 s.
    } \label{fig:lcprompt}
\end{figure}

The gamma-ray burst GRB\,000210 was detected simultaneously by the
{\em BeppoSAX} Gamma-Ray Burst Monitor (GRBM) and Wide Field
Camera 1 (WFC) on 2000 February 10, 08:44:06 UT. As of now, this
event is the brightest GRB detected simultaneously by the GRBM and
WFC, with a peak flux F(40-700 keV)=$(2.1\pm0.2)\times 10^{-5}$
erg cm$^{-2}$ s$^{-1}$, ranking in the top 1\% of the BATSE
catalog. In X-rays the event was also very bright, with a peak
flux F(2-10 keV)=$(1.5\pm0.2)\times10^{-7}$ erg cm$^{-2}$
s$^{-1}$, ranking fourth after GRB\,990712 \citep{fav+01},
GRB\,011121(Piro \etal, in preparation) and
GRB\,01022\citep{iz+01}. The gamma-ray light curve shows a single,
FRED-like\footnote{Fast Rise Exponential Decay} pulse
(Fig.~\ref{fig:lcprompt}), with a duration of about 15 s. The
X-ray light curve shows a longer pulse, with a tail persisting for
several tens of seconds. The fluence (40-700 keV)=$
(6.1\pm0.2)\times10^{-5}$ erg cm$^{-2}$ ranks GRB\,000210 in the
top 5 brightest GRBs seen by the GRBM and WFC, and in the top 3\%
of the BATSE bursts \citep{k+00}. With an F(2-10 keV)/F(40-700
keV)=0.007, GRB\,000210 is one of the hardest GRBs detected by
{\em BeppoSAX} \citep{fas+01}. The spectrum is also very hard  in
the X-ray band. Time resolved spectra of the WFC, fitted with a
power law model $F=C e^{-\sigma N_H}E^{-\Gamma}$, give
$\Gamma=(0.38\pm0.13)$ and $\Gamma=(0.82\pm0.12)$ in the rising
and first decaying part of the peak, with the usual hard-to-soft
evolution continuing in the subsequent parts, with
$\Gamma=2.3\pm0.15$ for $ 18s<t<80s$ (Fig.~\ref{fig:lcprompt}).
The absorption column density is consistent with that in our
Galaxy ($N_{HG}=2.5\times 10^{20}$ cm$^{-2}$), with an upper limit
$N_H\ltsim2\times 10^{22}$cm$^{-2}$.

\begin{figure}
\epsscale{0.9}
%\plotone{sax+cxo_bellaima.ps}
\plotone{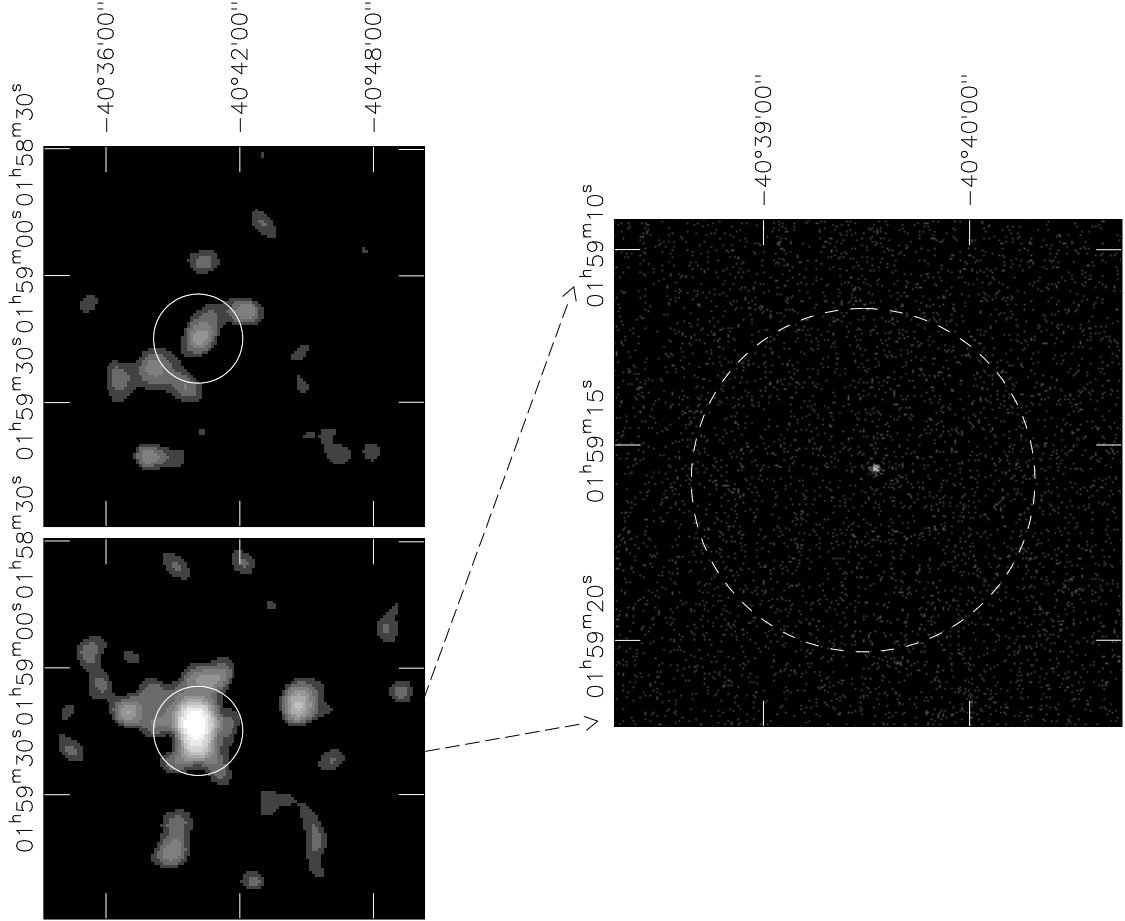}
%\plotone{test.ps}
\caption{Image of the X-ray afterglow of GRB\,000210 by {\em BeppoSAX} and
  {\em Chandra}. Upper panel: the left and right images show the
  afterglow in the {\em BeppoSAX} MECS(1.6-10 keV) 8 hrs and 30 hrs
  after the GRB respectively. The
  circle represents the WFC error box. Lower panel: the {\em Chandra}
  ACIS-S(0.2-8 keV)
  image of the afterglow 21 hrs after the GRB. The dashed circle is
  the {\em BeppoSAX} NFI error box} \label{fig:ximage}
\end{figure}

\begin{figure}
\epsscale{0.9}
%\plotone{gb000210_lc_final.ps}
\plotone{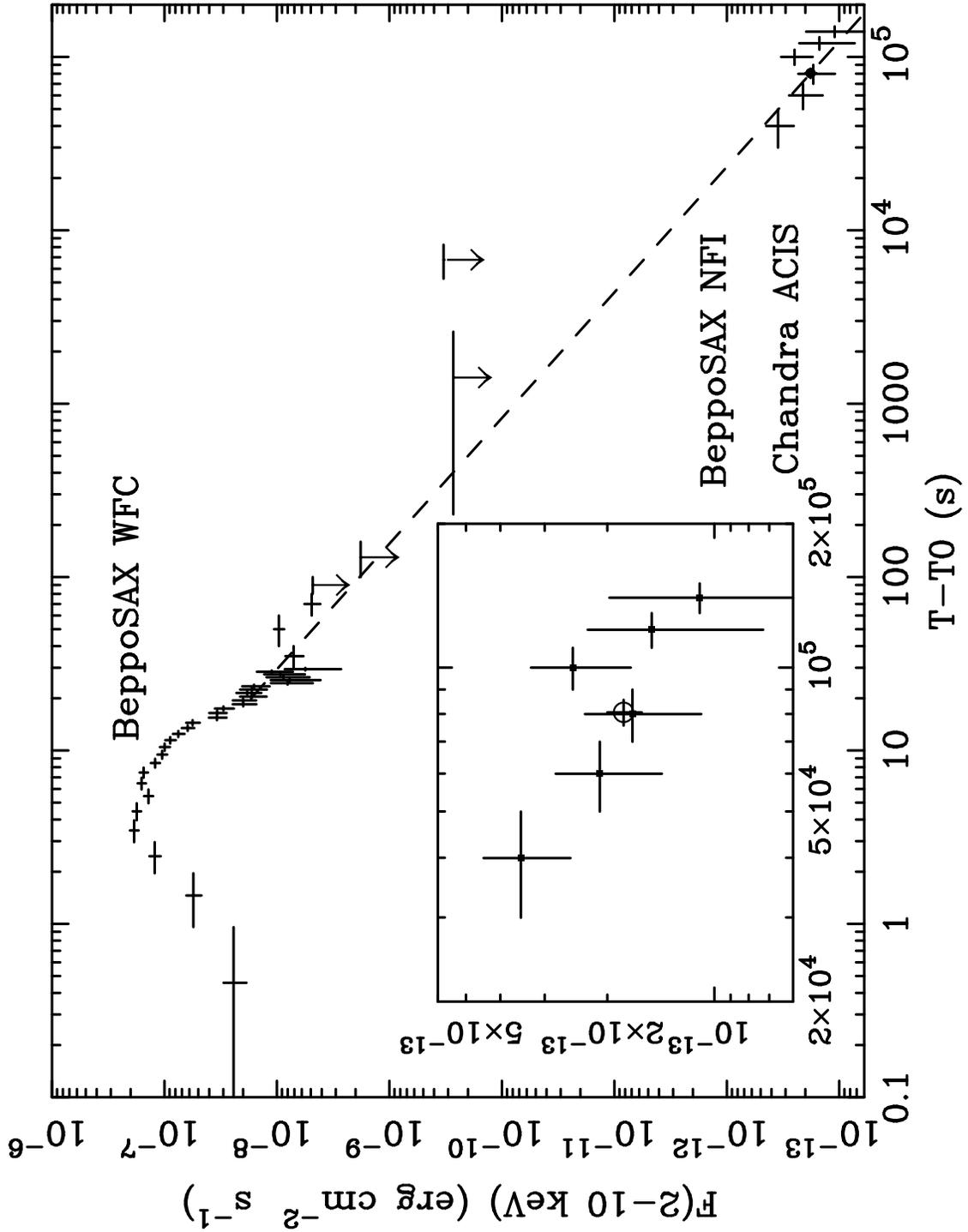}
  \caption{ Light curve from {\em
      BeppoSAX} WFC (from 0.1 to 5000 s), and {\em BeppoSAX} NFI and
    {\em Chandra} ACIS (from 30.000 to 140.000 s). The latter data are
    expanded in the inset. The Chandra data point is identified by
    an open circle }\label{fig:lcall}
\end{figure}

 The GRB was localized with the WFC at (epoch
J2000) R.A.\ = \ra{01} {59} {14.9}, dec.\ = $-40^\circ
40.14$\arcmin\ within a radius of 2\arcmin. This position is
consistent with the IPN annulus derived by Ulysses, Konus and the
BeppoSAX/GRBM \citep{hfp+00}.  Prompt dissemination of the
coordinates \citep{g+00,scm+00}) triggered follow-up observations
by several ground-based and space observatories, including {\em
BeppoSAX} and {\em Chandra}. A {\em
  BeppoSAX} target-of-opportunity observation (TOO) started on
Feb.~10.66 UT (7.2 hrs after the GRB) and lasted until Feb.~11.98
UT. Net exposure times were 44 ks for the MECS and 15 ks for the
LECS. The X-ray fading afterglow was detected \citep{cgp+00}
within the WFC error circle at (epoch J2000) R.A.\ = \ra{01} {59}
{15.9}, dec.\ = \dec{-40} {39} {29} (error radius=50\arcsec, see
Fig.~\ref{fig:ximage}). The X-ray flux from the source exhibited a
decay consistent with the standard power-law behavior, with F(2-10
keV)=$3.5\times 10^{-13}~\flux$ in the first 30 ks of the
observation (Fig.~\ref{fig:lcall}).  The spectrum derived by
integrating over the entire observation is well fitted by a power
law with photon index $\Gamma=1.75\pm0.3$, column density
$N_H<4\times 10^{21}$ cm$^{-2}$, consistent with that in our
Galaxy and flux F(2-10 keV)=$2.2\times 10^{-13}~\flux$.

The {\em Chandra} TOO started approximately 21 hrs after the burst
({\em i.e.,} around the middle of the {\em BeppoSAX} TOO), with an
exposure time of 10 ks with ACIS-S with no gratings. The X-ray
afterglow was near the center of the {\em BeppoSAX} NFI region
(Fig.~\ref{fig:ximage}). The initial position \citep{gpg+00}
derived from the first data processing by the Chandra X-ray Center
was found to be affected by an aspect error of about 8\arcsec\
\citep{ggp00}. Re-processing of the data with the correct attitude
calibration solved this problem \citep{gps+00}. We have further
improved this position, following the prescription suggested by
the Chandra team\footnote{http://asc.harvard.edu/mta/ASPECT/},
using 5 stars detected in X-rays and comparing their positions to
the USNO A2 and 2MASS catalog positions. The final position of the
GRB is (epoch J2000) R.A.\ = \ra{01} {59} {15.58}, dec.\ =
\dec{-40} {39} {33.02}, with an estimated error radius of
$0\farcs6$ (90\% confidence level).

\begin{figure}
\epsscale{0.8}
%\plotone{xspettro_nhvsz.ps}
\plotone{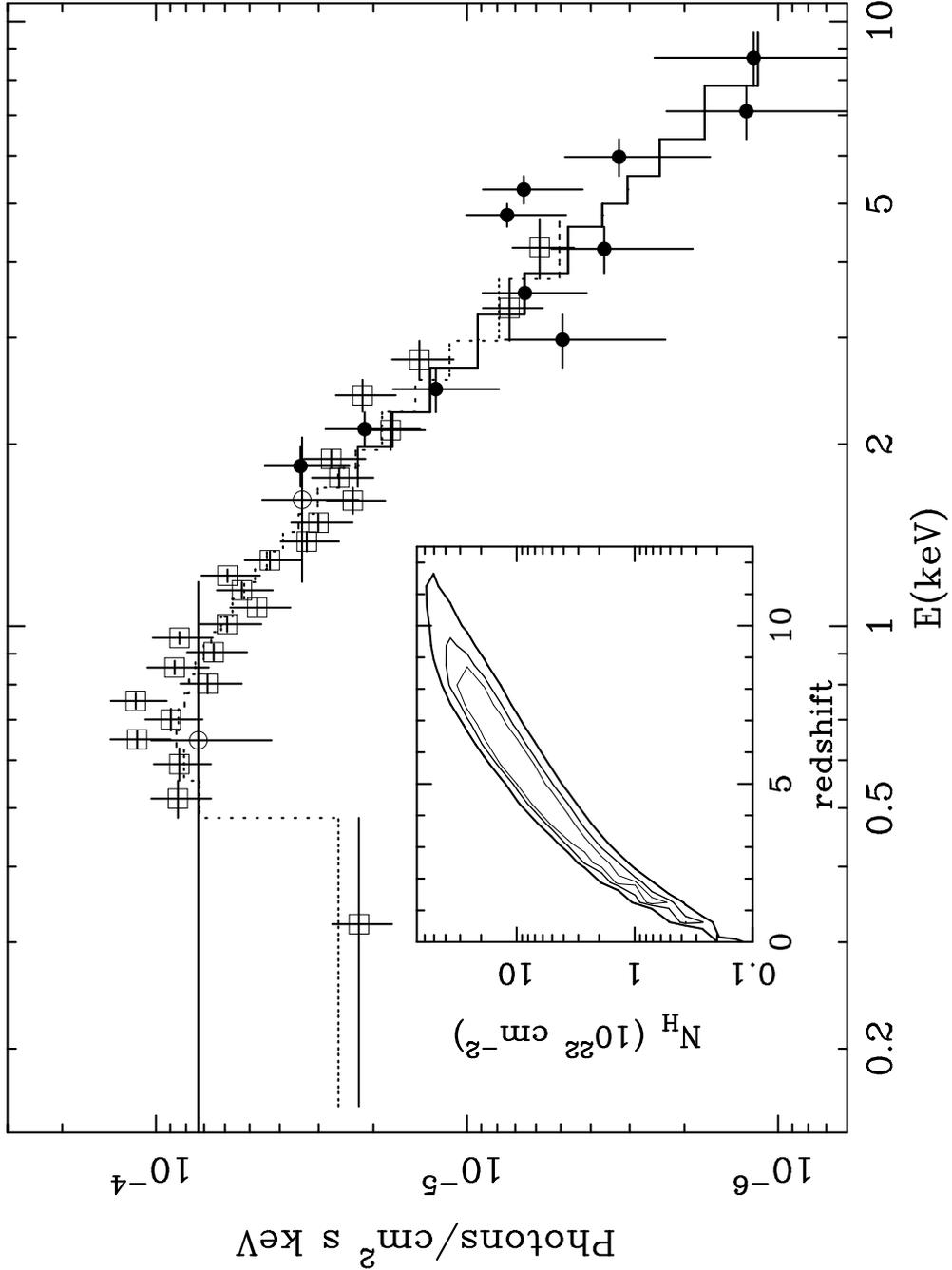}
  \caption{The X-ray spectrum of the
    afterglow by {\em BeppoSAX} LECS (empty circles), MECS (filled
    circles) \& {\em Chandra} ACIS-S (empty squares). The continuous
    (dashed) line is the best fit absorbed power law to the {\em
      BeppoSAX} ({\em Chandra}) data.  The contour plot of the intrinsic
      absorption column density as a function of the redshift is
      plotted in the inset. The contours correspond to 68\%, 90\% and 99\%
      confidence level (thin, normal, thick lines)} \label{fig:xspettro}
\end{figure}

The ACIS-S total counts were $555\pm26$, corresponding to F(2-10\
keV)=$1.8\times 10^{-13}~\flux$ for a best-fit power law
($\csq=20.9$ with 22 d.o.f.) with index $\Gamma=2.0\pm0.2$ and a
significant amount of absorption $N_{H}=(0.17\pm0.04)\times
10^{22}\rcm^{-2}$, well above that expected from our Galaxy.
%We note that the {\em Chandra} flux is
%perfectly consistent with the {\em BeppoSAX} data
%(Fig.~\ref{fig:lcall}).
%The temporal slope from BeppoSAX NFI and Chandra
%$F\propto t^{-\delta}$ is $\delta= 0.7\pm0.7$,
%$ F(t=1sec)=5.3\times 10^{-10}\flux$
We have performed a simultaneous fit to the {\em Chandra} ACIS-S,
and {\em BeppoSAX} MECS and LECS data with an absorbed power law,
leaving the relative normalization of the instruments free, to
account for the non-simultaneous coverage of the decaying source.
The resulting fit (Fig.~\ref{fig:xspettro}) is satisfactory
($\csq/{d.o.f.}=26/34$).  In particular the ACIS-S/MECS relative
normalization is $1.08\pm0.2$, and $\Gamma=1.95\pm0.15$. In the
inset of Fig.~\ref{fig:xspettro} we present the confidence levels
of the intrinsic hydrogen column density as a function of the
redshift of the source.

In Fig.~\ref{fig:lcall} we plot the light curve in the 2-10 keV
range from the prompt emission to the afterglow.  Frontera \etal\
(2000)\nocite{fac+00} have argued that {\it on average}, the
afterglow starts at a time $\approx 60\%$ of the duration of the
burst. This is consistent with what we observe in \grb. The steep
gradient of the WFC light curve flattens out around t=20 s,
suggesting that at this time the afterglow starts dominating over
the prompt emission. In fact, the WFC data points at $t\gtrsim 20
s$ and the {\em BeppoSAX} and {\em
  Chandra} data follow a  power law decay (F$\propto
t^{-\alpha_x}$) with $\alpha_x=1.38\pm0.03$ and F$(2-10 keV, t=11$
hrs$)=4\times 10^{-13}~\flux$. This interpretation is also
supported by the spectral behaviour, with the WFC spectral index
attaining a value of $\Gamma=(2.3\pm0.15)$ around t=20 s
consistent with that observed at later times by the {\em BeppoSAX}
NFI and {\em Chandra} ACIS-S.
%This decay slope is consistent with that derived from
%BeppoSAX/MECS and Chandra data only, equal to $-(0.8\pm0.6)$.
There is no evidence for a break in the X-ray light curve, which
would have been clearly detected if present in such a bright burst
(e.g. GRB\,990510, \citet{psa+01}, GRB\,010222, \citet{iz+01}).

\subsection{Optical Observations}\label{sec:optical}

We obtained optical imaging  of the field of \grb  ~with the
2.56-m Nordic Optical Telescope (NOT) equipped with the HIRAC
instrument and with the 1.54-m Danish Telescope (1.54D) plus DFOSC
starting 12.4 hours and 16 hours after the burst, respectively.
Further observations were carried out in August and in October
2000.  A log of these observations can be found in Table
\ref{tab:Table-Optical}, while in Table~\ref{tab:secondary} we
list the secondary standards used for the calibration of the
optical magnitudes of the field.

%We obtained optical observations of the field of GRB\,000210 with
%the 2.56m Nordic Optical Telescope (NOT), the 1.54D, the ESO NTT
%and 2.2 m from 12.4 hours to several months after the burst (see
%Tab for the observing log. ....

\begin{figure}
\epsscale{0.9} \plotone{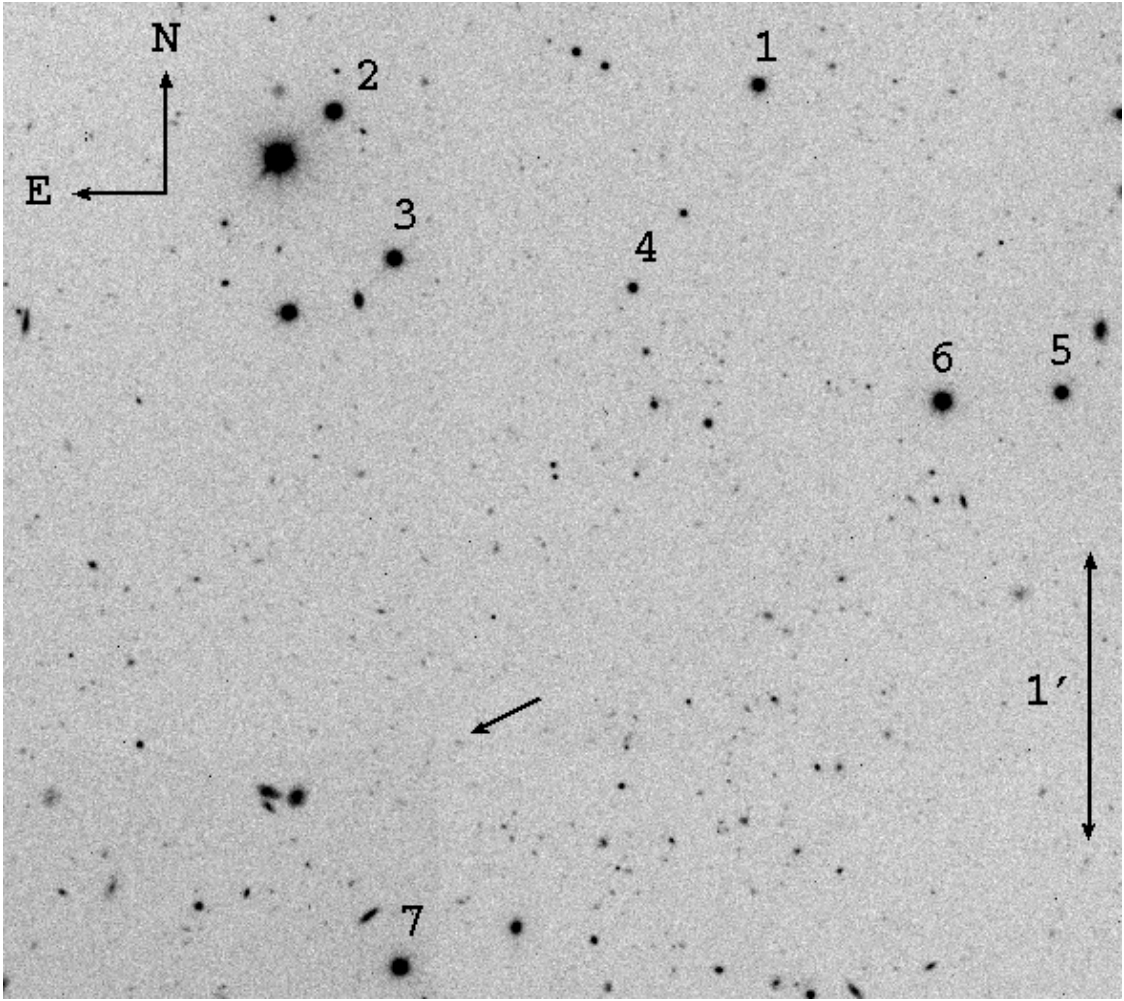}
  %\plotone{fig5.new2.ps}
  %\psfig{figure=chart8.blowup.ps, width=5cm}
  \caption{R-band VLT  image   of the GRB000210 field. The
  numbers label the secondary standards (Tab.~\ref{tab:secondary}) }
  \label{fig:optical}
\end{figure}

\begin{figure}
  %\plotone{chart9.blowup.ps}
  \epsscale{0.95}
  \plotone{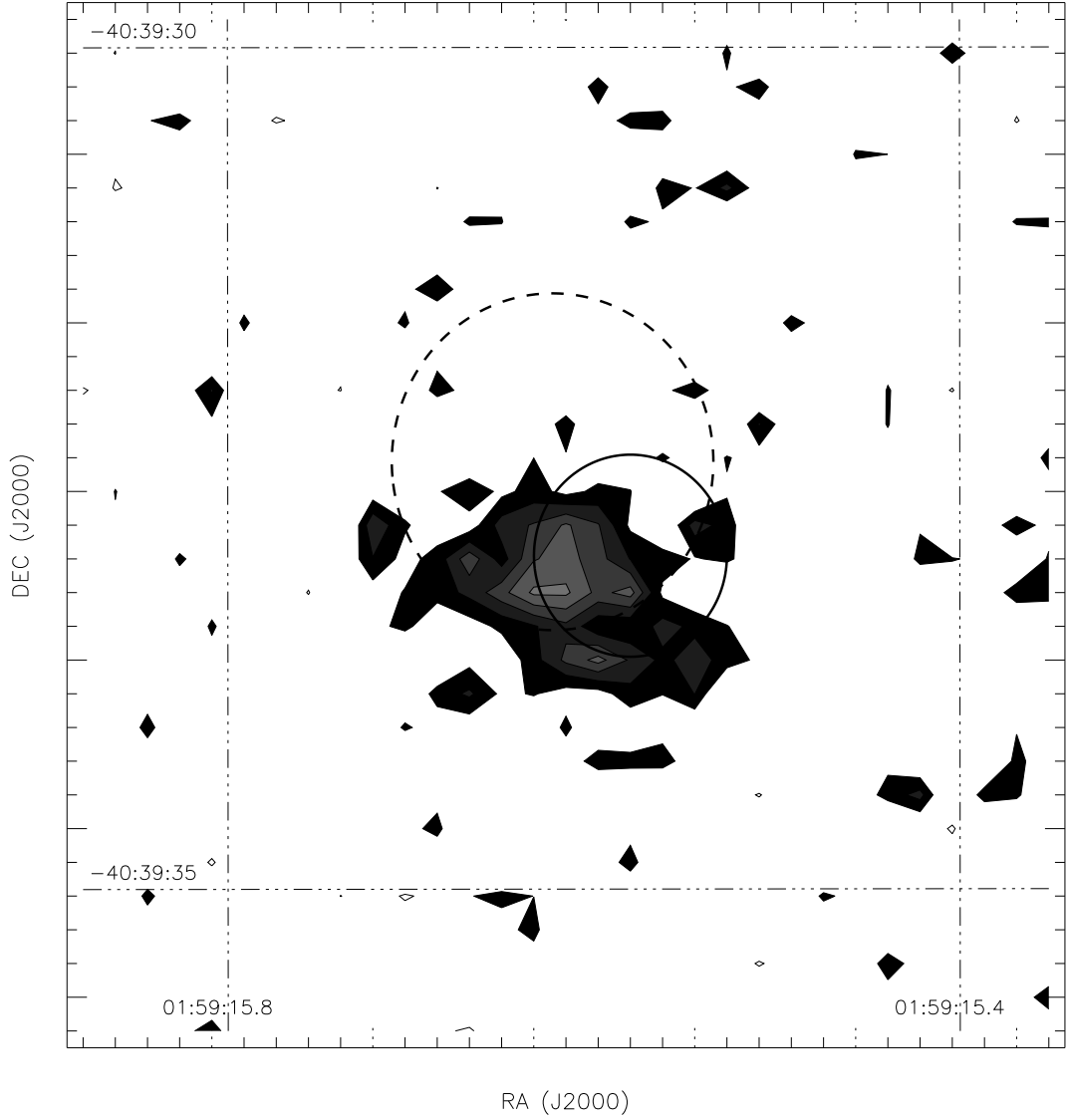}
  %\plotone{1sigma.2circles.new.ps}
  %\psfig{figure=chart8.blowup.ps, width=5cm}
  \caption{Blow-up of previous figure, showing the contour plot of
  the likely host galaxy of GRB000210. The circles show the 90\% error circles
   of the
  {\em Chandra} (continuous line) and radio (dashed line) afterglows.
  } \label{fig:oimage2}
\end{figure}

The observations (Fig.~\ref{fig:optical}) revealed a faint
extended object located within the {\em Chandra} error circle
\citep{gjo+00}. The contours of the optical emission
(Fig.~\ref{fig:oimage2}) show that the source is slightly
elongated in the North-East direction with an angular extension of
$\sim1\farcs5$. The angular size in the orthogonal direction
(North-West) is limited by the seeing ($0\farcs7$ in our best
images). The center of the object (i.e. excluding the diffuse
component) has coordinates (epoch 2000) R.A.\ = \ra{01} {59}
{15.61}, dec.\ = \dec{-40} {39} {33.1} with a 90 \% uncertainty of
$0\farcs74$ which represents the sum in quadrature of  a
systematic error of $0\farcs4$ and a statistical error of
$0\farcs6$. This position is the mean value of astrometric
calibrations on three independent images based on the USNO A2.0
catalog.  The position of two of the stars detected in the Chandra
image, that are included in the smaller field of view of the
optical image,  demonstrates that the optical and X-ray fields are
tied to within $\approx 0\farcs2$ (90\% confidence level). A
comparison of early and late observations shows that the object
remains constant in brightness within the photometric errors, with
magnitudes B=$25.1\pm0.7$, V=$24.1\pm0.15$, R = 23.5 $\pm$ 0.1,
I=$22.60\pm0.12$. Although the object in Fig.~\ref{fig:oimage2} is
very faint, its appearance is not stellar.

\subsection{The host galaxy ?}

Spectroscopic observations confirmed that the object is a galaxy.
The observations were carried out on 2000 October 25 with the VLT1
equipped with FORS1. The 300V grism and the $0\farcs7$ slit
provided a spectral FWHM resolution of $\sim10$ \AA.  The spectrum
is based on a single 2000 sec exposure and covers the range
between $4193$ \AA\ and $8380$ \AA.  The reduction is based on
standard procedures, {\em
  i.e.}, flat field correction with internal lamp flats and wavelength
calibration using arc lamps.  No spectro-photometric standard
stars were observed, so no flux calibration was possible.

\begin{figure}
%\plotone{spectrum.1.ps}
\plotone{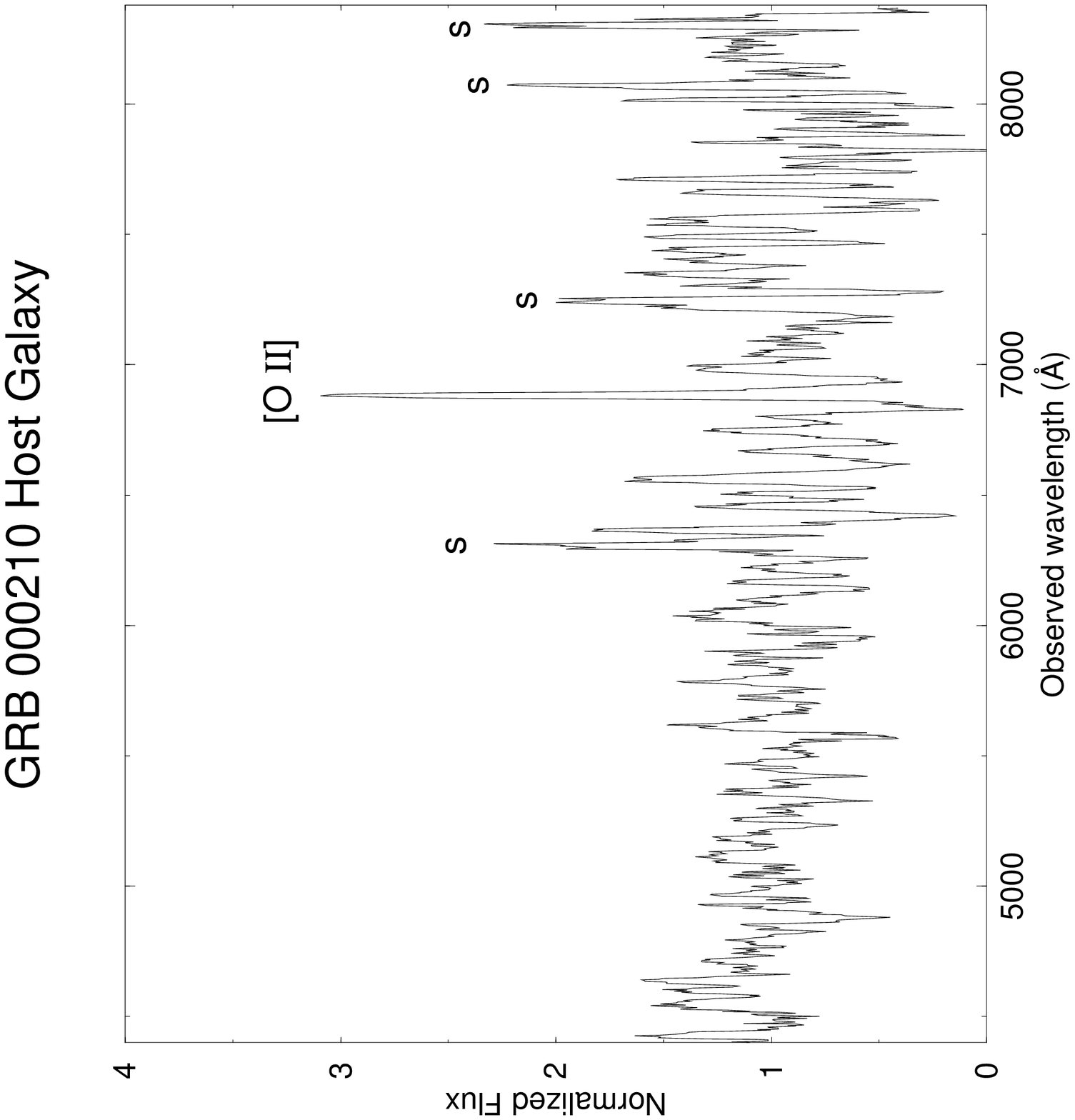} \caption{
%{\em top panel:}
The  normalized spectrum of the GRB 000210 host galaxy acquired
with VLT+FORS1.  The spectrum has been smoothed with a boxcar
width corresponding to the instrumental spectral  resolution (10
\AA). The spikes, mostly present  in  the red part  of the
spectrum and indicated with  an S, are  residuals  from the
subtraction  of strong sky emission lines. The region of the [OII]
line is not affected by sky emission lines.
% {\em
%bottom panel:} The figure shows the  sky spectrum  as  well as the
%extracted trace of  the  object, measured in CCD counts per
%second.  As can be seen the lines indicated with an S in the upper
%panel correspond to  regions dominated by strong emission lines.
%The [OII] line is located in a gap between sky emission lines.
} \label{fig:ospectrum}
\end{figure}

The spectrum revealed an emission line at $6881.2\pm$0.5 \AA,
$\sim6\sigma$ above the continuum level
(Fig.~\ref{fig:ospectrum})). Given the presence of a well-detected
continuum blueward of the line, and the absence of other emission
lines in the $4193-8380$ \AA~ range, we identify this line with
the $3727$ \AA [OII] line at a redshift of $z=0.8463 \pm 0.0002$.
The equivalent width of the line is $EW=(68\pm9)$ \AA. We stress
that the absence of V dropout in photometric data, expected by
Ly$_\alpha$ forest absorption at high z, indicates by itself that
$z<4$ \citep{mad95}.

Is this galaxy the {\it host} of GRB\,000210?  We have computed
the probability of a chance association in two ways. First we have
computed the probability of finding an unrelated field galaxy with
$R<R_{host}=23.5$ within the localization circle  of the afterglow
( $1^{\prime \prime}$ radius, i.e. the 99\% {\em Chandra} error
radius). From galaxy counts \citep{hogg+97} we derive that $P=
10^{-2}$. A slightly more conservative estimation is given by the
the fraction of the sky covered by galaxies brighter than
$R_{host}$:

\begin{equation}
P=\int^{R_{host}}{A(m)\frac{dN}{dm} dm} \label{eq:p1}
\end{equation}
where $dN/dm$ is  the mean number of galaxies per magnitude per
unit solid angle and $A(m)$ is the average area of a galaxy  of
R-band magnitude $m$. For $m>21$ $dN/dm\approx 10^{0.334*m}$
\citep{hogg+97}, while for brighter galaxies
$dN/dm\approx10^{0.5*m}$ \citep{kk92}. We have estimated the
average area of a galaxy $A=\pi (2r_{hl})^2$, where $r_{hl}$ is
the half-light radius of the galaxy. We have adopted the following
empirical half-light radius-magnitude relations:
$r_{hl}=0\farcs6\times10^{-0.075*(m-21)}$, for $21<m<27$
\citep{owd+96,bkd02}, and $r_{hl}=0\farcs6\times10^{-0.2*(m-21)}$
for brighter galaxies \citep[][and references therein]{icg+95}. By
substituting these relations in eq.\ref{eq:p1}, we derive:

\begin{equation}
P=[4.9+3.8\times10^{0.184(R_{host}-21)}]\times10^{-3}\quad
21<R_{host}<27
\label{eq:p2}
\end{equation}

For $R_{host}=23.5$, $P=1.6\times 10^{-2}$. Therefore, a chance
association is unlikely, but not completely negligible.
%Therefore, if this object is the host galaxy of GRB\,000210, we are
%brought to the conclusion that the lack of an optical transient in
%{\it this} GRB, is not due to the Ly absorption

\subsection{Radio Observations}\label{sec:radio}

Very Large Array (VLA)\footnotemark\footnotetext{The NRAO is a
  facility of the National Science Foundation operated under
  cooperative agreement by Associated Universities, Inc.  NRAO
  operates the VLA.} observations were initiated within 15 hours after
the burst. Details of this and all subsequent VLA observations are
given in Table \ref{tab:Table-Radio}. In addition, a single
observation was also made two days after the burst with the Australian
Telescope Compact Array (ATCA).\footnotemark\footnotetext{The
  Australia Telescope is funded by the Commonwealth of Australia for
  operation as a National Facility managed by CSIRO.} All VLA
observations were performed in standard continuum mode, with a central
frequency of 8.46 GHz, using the full 100 MHz bandwidth obtained in
two adjacent 50 MHz bands. The flux density scale was tied to the
extragalactic source 3C48 (J0137+331), while the phase was monitored
using the nearby source J0155$-$408. The ATCA observation was made at
a central frequency of 8.7 GHz, with a 256 MHz bandwidth obtained in
two adjacent 128 MHz bands. The flux density scale was tied to the
extragalactic source J1934$-$638, while the phase was monitored using
the nearby source J0153$-$410.

On 2000 February 18.95 UT a radio source was detected
%(Fig.~\ref{fig:radio})
within the {\em Chandra} error circle with a peak brightness of
99$\pm$21 $\mu$Jy beam$^{-1}$, or a flux density (after Gaussian
fitting) of 93$\pm$21 $\mu$Jy. The synthesized beam was 6\arcsec\
by 2\arcsec, with a position angle on the sky of 30$^\circ$
counterclockwise. The position of this source is (epoch J2000)
R.A.\ =\ra{01} {59} {15.57} ($\pm{0.05^s}$), dec.\ = \dec {-40}
{39} {31.9} ($\pm1\farcs0$), where the errors are at 90 \%
confidence level in the Gaussian fit. We have further refined the
astrometry, by looking for coincident USNO2 stars. We find one
radio source coincident with a star, shifted by an offset of
($-$0.05s, $0\farcs56$) in R.A. and dec. Taking into account this
offset, we derive the final position of the radio counterpart of
the GRB at R.A.\ =\ra{01} {59} {15.62}, dec.\ = \dec {-40} {39}
{32.46}. The position of this radio transient is in excellent
agreement with the X-ray afterglow and the optical source
(Fig.\ref{fig:oimage2}). The transient nature of this source is
readily apparent (see Table \ref{tab:Table-Radio}), since it was
not detected in observations either before or after February 18.

\section{Discussion}\label{sec:discuss}

\subsection{ Properties of the host}

In the following we will assume a cosmology with H$_{\circ}=$65 km
s$^{-1}$ Mpc$^{-1}$, $\Omega_m=0.3$, and $\Omega_{\Lambda}=0.7$.
At $z=0.846$, the luminosity distance $D_L=1.79\times 10^{28}\
$cm, and 1 arcsec corresponds to 8.3 proper kpc in projection. The
$\gamma$-ray fluence implies an isotropic $\gamma$-ray energy
release $E_{\gamma}=1.3\times 10^{53}\ $ erg.

The [OII]3727 line flux can be used to estimate the star formation
rate of the galaxy \citep{ken98}. Although the spectrum has not
been calibrated, we have derived a rough estimation of the line
flux by rescaling the line flux measured in the host galaxy of
GRB\,970828 for the line EW, R magnitudes and redshifts
\citep{dfk+01}. We find $L_{3727}\approx 1-2\times 10^{41}$ erg
s$^{-1}$, corresponding to SFR$\approx 3 \Msun\ $ yr$^{-1}$. Since
we are not sensitive to any obscured components of SFR, this
relatively modest SFR is only a lower limit to the true star
forming rate. From the I band photometry we derive a rest frame
absolute magnitude M$_B=-19.9\pm0.1$, corresponding to a $\approx
0.5 L_*$ galaxy today \citep{schec76}. The location of the
afterglow is within $\approx$1\arcsec\ the center of the galaxy,
corresponding to about 8 kpc in projection.

\subsection{The nature of the obscuring medium}

The {\em Chandra} and {\em BeppoSAX} observations detected an
X-ray afterglow which clearly underwent a power-law decay with
$\alpha_x=1.38\pm 0.03$.  There is no evidence for a temporal
break in the X-ray light curve between 10 s and 2$\times 10^{5}$ s
after the burst (Fig.~\ref{fig:lcall}).  A deviation from a power
law decay could occur if the synchrotron cooling break $\nu_c$
passed through the band, or the outflow began to exhibit jet
geometry, or at the transition to non-relativistic expansion. In
all respects the X-ray afterglow of \grb\ was fairly typical in
comparison with past events \citep{piro01}.

The VLA and ATCA measurements indicate that the only significant
detection of the radio afterglow from \grb\ was nine days after
the burst. With a 8.46 GHz flux density of 93$\pm$21 $\mu$Jy, this
is the weakest radio afterglow detected to date. The time interval
between the burst and the radio peak is too long to be the result
of a reverse shock propagating in the relativistic ejecta
\citep{kfs+99,sp99a}. It is more likely that the emission
originated from a forward shock which reached a maximum on this
timescale  \citep[e.g.][]{fkb+99}. Interstellar scintillation can
briefly increase (or decrease) the radio flux of a weak afterglow
and make it detectable \citep{goo97,wkf98}. With these
single-frequency measurements it cannot be determined whether the
8.46 GHz flux density was weak because \grb\ was a low energy
event, or because the synchrotron self-absorption frequency was
high ($\nu_{ab}>10$ GHz). Reichart \& Yost (2001)\nocite{ry01}
have argued that a high $\nu_{ab}$ is expected for dark bursts if
they occur in dense circumburst environments ($n>>10^{2}$
cm$^{-3}$).

Although the X-ray and radio afterglow were detected for \grb,
there are only lower limits for the magnitude of the expected
optical transient of R$>22$ and R$>23.5$ at 12.4 hrs and 16 hrs
after the burst, respectively.
%This absence cannot be readily
%explained by a failure to image the GRB field quickly enough or
%deeply enough since almost all of the {\it detected} optical
%afterglows have R$<$24, 18 hrs after a burst \citep{lcg01,ry01}.
Following Djorgovski {\em et al.} (2001) we can use the fireball
model to predict the expected optical flux and then derive a lower
limit to the amount of the extinction. The simplest model of an
isotropic fireball expanding into a constant density medium is
assumed \citep{spn98}. This assumption is well justified since all
well studied afterglows are better explained by expansion in a
constant density medium rather than in a wind-shaped one
\citep{pk01}. For the timescales of interest ($t<21$ hrs), there
is no evidence for a break in the X-ray light curve that would
indicate a jet-like geometry or transition to non-relativistic
expansion. Two limiting cases are considered. The first is when
the cooling frequency $\nu_c$ lies below the optical band $\nu_o$
({\em i.e.,} $\nu_c<\nu_o$). The expected optical spectral flux
density is $f_o=f_x(\nu_o/\nu_x)^{-p/2}$, where $f_x$ is the X-ray
spectral flux density at $\nu_x$. The second case is where the
cooling frequency lies between the optical and X-ray bands ({\em
i.e.,} $\nu_o< \nu_c < \nu_x$). There is a local minimum when
$\nu_x\simeq\nu_c$ and consequently the expected optical spectral
density is $f_o=f_x(\nu_o/\nu_x)^{-(p-1)/2}$. As long as $\nu_c
\leq \nu_x$, the electron energy spectral index
p=($2/3+4/3\alpha_x$)=2.51$\pm$0.04. This value of p is also
consistent with the spectral slope measured in X-rays.

The X-ray flux measured by {\em Chandra} 21 hrs after the burst
was F(2-10 keV)=1.8$\times 10^{-13}$ erg cm$^{-2}$ s$^{-1}$, which
corresponds to a spectral flux density $f_x=0.01$ $\mu$Jy at 4 keV
(adopting $\beta_x=(\Gamma-1)=0.95\pm{0.15}$ where f$_x\propto
\nu^{-\beta_x}$). For $\nu_c<\nu_o$ the predicted optical
magnitudes are therefore R$\sim$17.4 and R$\sim$ 17.7 at 12.4 hrs
and 16 hrs after the burst, respectively. The equivalent
magnitudes for the case $\nu_o< \nu_c \simeq \nu_x$ are
R$\sim$21.5 and R$\sim$21.9. Taking the more stringent limits on
the absence of an optical transient 21 hrs after the burst, we
infer significant extinction towards \grb, with a range of upper
limits lying between A$_{\rm R}$=1.6 and 5.8 mag.  Using the
extinction curve as formulated by Reichart (2001)\nocite{rei01},
we convert these A$_{\rm R}$ limits from the observer frame to a
rest frame extinction of A$_{\rm V}$=0.9 to 3.2 (for $z=0.85$),
and derive a hydrogen equivalent column density $N_{HO}\gsim
0.2-0.6\times{10}^{22}$ cm$^{-2}$ assuming the Galactic relation
of Predehl \& Schmitt (1995)\nocite{ps95}.

How does this result compare with the estimate of absorption
derived from the X-ray data?  In Fig.~\ref{fig:xspettro} we show
the contour plot of the X-ray column density in the GRB frame as a
function of the redshift, under the assumption that the absorbing
material is in a neutral cold state. At $z$=0.85 the absorption is
$N_{HX}=(0.5\pm0.1)\times 10^{22}~\ha$. Thus $N_{HX}/N_{HO}$ is
consistent with unity, or less, and therefore the dust-to-gas
ratio is compatible with that of our Galaxy, as it has also been
found in the other dark GRB\,970828 \citep{dfk+01}. In contrast,
Galama \& Wijers (2000)\nocite{gw01} noted that a number of bursts
with optical afterglows seem to exhibit large column density as
inferred from X-ray afterglow data, but with little or no optical
extinction, suggesting that GRBs destroy dust grains along the
line of sight. Different authors \citep{wd00,fkr01,rei01c} present
mechanisms by which dust in the circumburst medium is destroyed or
depleted by the light from the optical flash and X-rays from the
burst and early-time afterglow up to a distance $R\approx10
L_{49}^{1/2}$ pc (L=$10^{49}L_{49}$ erg s $^{-1}$ is the
isotropic-equivalent luminosity of the optical flash). Reichart
(2001) \nocite{rei01c} and Reichart \& Price (2001) \nocite{rp01}
proposed a unified scenario to explain the two populations of
GRBs. They argue that most  GRB occur in giant molecular clouds
(GMC), with properties similar to those observed in our Galaxy
(size $\approx20-90 $pc, \citet{srb+87}). Assuming that the energy
reservoir is standard \citep{fks+01,pkpp01}, strongly collimated
bursts would burn out completely through the clouds, thus
producing a detectable optical afterglow regardless of the column
density through the cloud. Weakly collimated bursts would not
destroy all the dust, leaving a residual column density through
the line of sight. If this column density is sufficiently high,
optical photons of the afterglow will be extinguished, making a
dark GRB.

We find, however, that this scenario is not consistent with the
observed properties of the X-ray absorber, if GRB\,000210 lies at
z=0.846. For typical densities of a GMC ($n\approx
10^2-10^5$~cm$^{-3}$), the gas should be ionized by GRB photons on
scales of several parsecs \citep{bdcl99,lp02}. Over the whole
cloud, the medium would span a wide range of ionization stages,
from fully ionized to neutral, with a substantial fraction of the
gas being in partially ionized stages. We have verified this case
by fitting the {\em BeppoSAX} and {\em Chandra} spectra with an
ionized absorber (model {\it absori} in XSPEC) at redshift
$z$=0.85. The ionization stage is described by the ionization
parameter $\xi=L_X/nR^2$, where R is the distance of the gas from
the GRB. We do not find any evidence of an ionized absorber, with
a tight upper limit of $\xi<1$. In fact, even a moderate level of
ionization would result in a reduced opacity below $\approx 1 $
keV, due to the ionization of light elements such as C and O, but
this feature is not observed in the X-ray spectrum.

\subsubsection{High density clouds}
We have shown above that in the GMC scenario the X-ray absorbing gas
is expected to be substantially ionized, contrary to what is observed.
We now introduce a variation of the {\it local absorption} scenario,
in which a phase of the medium is condensed in high density clouds or
filaments with low filling factor. We show that this scenario is
consistent with the properties of the X-ray absorption.

To keep the gas in a low-ionized stage at a distance of the order
of few pc,  a density $n\gtrsim10^{9}$cm$^{-3}$ is required. In
fact, the recombination time scale $t_{rec}\approx
n^{-1}T^{-0.5}$, where $T$ is the electron temperature. In the
case of iron, $t_{rec}=300 n_9^{-1}T_7^{-0.5} s$ ( {\em i.e.}
\citet{pgg+00}; hereafter, given a quantity $X$, we define
$X_n=10^{-n} X$). The typical temperature is expected to be in the
range $T_7\approx0.1-1$ \citep{pgg+00,pkhl00}. Therefore, at
sufficiently high density, recombination is effective in keeping
the gas close to ionization equilibrium over timescales $\gtrsim
t_{rec}$, {\em i.e.}, during the afterglow phase. The ionization
parameter is then $\xi=\frac{L_{45}}{n_9 R_{18}^2}$, where
$L_{45}=3$ is the luminosity of the X-ray afterglow in the rest
frame energy range 0.013-100 keV.  For $n_9\gtrsim 1$ the medium
is then in the neutral phase for distances $R_{18}\gtrsim{1}$, as
required. It is straightforward to show that this medium should be
clumpy. The size of each clump has to be $r_s\ltsim N_H/n \ltsim
5\times10^{12}/n_9$ cm, and the fraction of volume occupied by
this medium ({\em i.e.}, the filling factor) $f_V=N_H/(nR)=\frac{5
\times 10^{-6}}{n_9 R_{18}}$. Finally we note that the size of a
clump is much smaller than the zone of the fireball visible at the
time of the observation, $\approx 10^{15}/\Gamma_b$, where
$\Gamma_b\approx2-10$ is the Lorentz factor of the fireball in the
afterglow phase. We therefore require that most of the source is
covered by these clouds, {\em i.e.}, that the covering fraction
$f_{cov}=f_V \frac{R}{r_c}\approx1$. This condition is satisfied
when $r_s \approx 5\times 10^{12}/n_9$ cm. The total mass
contained in these clouds is $M_c\approx 3R_{18}^3
f_{V,-6}n_9~\Msun$.

It is well known from observations and models
\citep{bwc01,wsha94}, that the medium in star--forming molecular
clouds is clumpy, with dense clouds or filaments embedded in a
much less dense intercloud medium. The largest densities are of
the order of $\approx 10^7$ cm$^{-3}$ \citep{nbh+00}, {\it i.e.}
lower than required. However, both observations and modeling are
limited in resolution to structures of size $\gtrsim10^{16}$cm and
would then miss smaller and higher density fluctuations.
Furthermore, the total mass of the X-ray absorbing clouds is a
tiny fraction ($\approx10^{-4}$) of the mass of a GMC ($\approx
10^{5-6} \Msun$).  {\it Prima facie} those structures could then
be the tail in the power spectrum of density fluctuations in star
forming regions.  Interestingly, Lamb (2001)\nocite{lamb01} has
also stressed the role of a {\it dusty} clumpy medium in GMC with
regard to optical properties of GRBs. A more detailed discussion
is beyond the scope of this paper.

It is worth noting that the density of this medium is similar to
that of the gas responsible for iron features
\citep[e.g.][]{pcf+99,pgg+00,afv+00}. This gas lies much closer to
the burst and is therefore highly ionized.  Interestingly, in the
afterglow of GRB000210, we find a very marginal evidence of a
recombination edge by H-like Fe atoms (at a confidence level of
$97\%$). The rest-frame energy of this feature is at 9.28 keV,
corresponding, at $z$=0.85, to E=5 keV, where {\em
  BeppoSAX} data show some residual (Fig.~\ref{fig:xspettro}).  The
$EW= (1\pm0.7) $ keV, is similar to that observed in GRB\,991216
\citep{pgg+00} and in the other dark GRB\,970828 \citep{yym+01}.
This feature is the result of electron recombination on H-like Fe
atoms and should be accompanied by a K$\alpha$ line at 6.9 keV
with a similar intensity \citep{pgg+00}. Within the errors, the
upper limit $EW<0.5\ $ keV to the latter line is consistent with
this prediction.

\subsubsection{ISM absorption in the host galaxy}

We now discuss the hypothesis that the absorption does not take
place in the circumburst environment of the GRB, but in the ISM of
the host galaxy. We find that this scenario can easily account for
the properties of this burst. In fact, it immediately explains the
absence of ionization features in the X-ray absorber and a
dust-to-gas ratio consistent with that in our Galaxy. The typical
column density through a GMC is about $10^{22}$ cm$^{-2}$
\citep{srb+87}, and therefore even a single GMC in the line of
sight could provide the necessary absorption both in the optical
and in X-rays. This scenario is also consistent with the location
of the GRB, which lies within $\approx 10$ kpc from the center of
the galaxy.

Could  non-local ISM absorption by the host galaxy be the unique
origin of the whole population of dark GRBs? Let us first assume
that GRBs occur in the disk of a galaxy similar to ours at a
typical distance of 10 kpc from the center. This assumption is
consistent with the {\it visible} properties of GRB host galaxies
and the distribution of GRB offsets with respect to the host
centre \citep{bkd02}. The line-of-sight column density of
interstellar hydrogen gas measured from our location in the Galaxy
(10 kpc from the center) is consistent with that required to make
a GRB dark ($N_H\gtrsim 5\times 10^{21}$cm$^{-2}$) in a belt of
about $\pm5\degr$ along the
 plane of the Galaxy \citep{dl90}.
Thus, of the entire population of  GRBs occurring in the disk of
galaxies randomly oriented in the sky, only   $\ltsim 10\%$ would
be dark, much below the observed fraction of $50-60\%$. Therefore,
while this hypothesis can account for a fraction of dark GRB's,
including this one,  in order to make up the entire population of
dark events, GRB host galaxies should contain quantities of dust
and gas, associated with {\it obscured} star-forming regions, much
larger than a typical galaxy like ours.

\subsubsection{Implications for the absorber in the high-z scenario }

Finally, given the modest {\it a posteriori} probability, we have
to consider the possibility that the association of GRB\,000210
with the galaxy is coincidental, and that this GRB is located at
$z\gtrsim5$. In this case the X-ray absorber of GRB\,000210 is
thicker and it can also be much more ionized.  For example, at
$z$=5, $\xi\ltsim10^3$, {\em i.e.}, the data are consistent with
an absorber from neutral to highly ionized. In particular we have
satisfactorily fitted the data either with a neutral absorber
($\xi=0$), that requires $N_H=9\times10^{22}$ cm$^{-2}$, or an
ionized absorber ($\xi=400$), that gives $N_H=16\times10^{22}$
cm$^{-2}$. Such  values of column densities are much higher than
those observed in X-ray afterglows of GRBs with optical transients
\citep[e.g.][]{pgg+01,iz+01}.

%We point out that we have no evidence so far of any other event
%with such property.
%  the point we want to
%address is whether the {\it high-z} scenario is sufficient to
%explain the difference between dark GRB's and those GRBs  (OTGRB).
%In Fig.~\ref{fig:ionized} we have shifted these model spectra  at
%$z$=1, {\em i.e.}, the average redshift of the population of
%OTGRBs. None of the spectra is consistent with the X-ray spectral
%properties of afterglows of OTGRBs. In the neutral case, we would
%expect a highly absorbed spectrum
%($N_{Hobs}\approx(1+z)^{-8/3}N_{Hrest}\approx 10^{22}$ cm$^{-2}$)
%with no emission below 1 keV. In the ionized case, a deep and
%broad trough between 0.4 and 1 keV should be present. None of
%these features is observed Therefore we conclude that GRB\,000210
%(and therefore dark GRBs, if GRB\,000210 is a typical
%representative of the class), if at $z\gtrsim5$, should be
%characterized by an intrinsically different X-ray afterglow
%spectrum compared to OTGRBs, with a much higher column density.

%-----

%Thence, it is well possible that the GRB lies
%in a dust lane or a spiral arm of the galaxy, as we have argued
%from the discussion on the absorption.

\section{Conclusions}

In this paper we have presented the results of multi-wavelength
observations of GRB\,000210. This event was the brightest ever
observed in $\gamma$-rays in the {\em BeppoSAX} GRBM and WFC.
Nonetheless, no optical counterpart was found down to a limit of
$R=23.5$.  GRB\,000210 is therefore one of the events classified
as dark GRBs, a class that makes up $\approx 50\%$ of all GRBs. It
is still unclear whether  this class derives from a single origin
or it is due to a combination of different causes. Some of these
GRBs could be intrinsically faint events, but this fraction cannot
be very high, because the majority of dark GRBs shows the presence
of an X-ray afterglow similar to that observed in GRBs with
optical afterglows \citep{piro01,lcg01}. The most compelling
hypotheses to explain the origin of dark bursts involve
absorption, occurring either in the local environment of the GRB
(circumburst or interstellar), or as Ly$\alpha$ forest absorption
for those bursts which have $z\gsim5$.

As in the majority of bursts, \grb\ had an X-ray afterglow which
was observed with {\it BeppoSAX} and {\it Chandra}.  The temporal
behavior is well described by a power law, with a decay index
$\alpha_x=-1.38\pm0.03$, similar to that observed in several other
events  \citep[e.g.][]{piro01}.  We did not find evidence for
breaks in the light curve. The spectral index of the power law is
also typical ($\Gamma=1.95\pm0.15$).

Thanks to the arcsecond localization provided by {\em Chandra} we
identified the likely host galaxy of this burst, determined its
redshift ($z$=0.846) and detected a radio afterglow. The
properties of the X-ray afterglow allowed us to determine the
amount of dust obscuration required to make the optical afterglow
undetectable ($A_R\gtrsim2$). The X-ray spectrum shows significant
evidence of absorption by neutral gas ($N_{HX}=(0.5\pm0.1)\times
10^{22}~\ha$). However, we do not find evidence of a partially
ionized absorber expected if the absorption takes place in a Giant
Molecular Cloud, as recently suggested to explain the properties
of the dark GRBs  \citep[e.g.][]{ry01}. We conclude that, if the
gas is local to the GRB, it has to be condensed in dense
($n\gtrsim10^9$ cm$^{-3}$) clouds. We propose that these clouds
represent the small-scale high-density fluctuations of the clumpy
medium of star-forming GMCs.

Both the amount of dust required to extinguish the optical flux
and the dust-to-gas ratio are consistent with those observed
across the plane of our Galaxy. We cannot therefore exclude that
the absorption takes place in the line-of-sight through  the
interstellar medium of the host, rather than being produced by a
GMC embedding the burst. This hypothesis is also consistent with
the location of  GRB\,000210 with respect to the center of the
likely host galaxy. To explain the whole population of dark GRB
this hypothesis would require that  host galaxies of GRBs should
be characterized by quantities of dust and gas  much larger than
typical,  arguing again for a physical connection between GRBs and
star forming regions.

Finally, we discussed the possibility that the galaxy is unrelated
to \grb\ and that it is a dark burst because it lies at
$z\gtrsim5$. In this case the X-ray absorbing medium should be
substantially thicker than that observed in GRBs with optical
afterglows.  Assuming that GRB\,000210 is a typical representative
of a population of events at high redshift, then these GRB's are
embedded in a much denser environment than that of closer events.

Whichever of the explanations apply, it is clear that dark GRBs
provide a powerful tool to probe their formation sites and
possibly to explore the process of star formation in the Universe.
We have at hand several observational tools to pursue this
investigation. By increasing the number of arcsecond locations by
radio, X-ray and far-infrared observations we can build up a
sample of host galaxies of dark GRBs and study their distances and
physical properties. The origin of the absorption in X-rays and
optical can be addressed by broad band spectra and modeling,
providing information on the dust and gas content of the absorbing
structures. X-ray measurements are particularly promising in this
respect for several reasons. First, X-rays do not suffer from
absorption, in fact roughly the same number of dark GRB and GRBs
with optical afterglows have an X-ray afterglow. Detection of
X-ray lines can thus provide a direct measurement of the redshift.
A comparative study of the X-ray properties of these two classes
should also underline differences that can be linked to their
origin, like the brightness of X-ray afterglows and the amount of
X-ray absorption. Finally, measurements of variability of the
X-ray absorbing gas would provide strong support to the {\it local
absorption} scenario. There are several mechanisms that can
produce such a variability. The hard photon flux from the GRB and
its afterglow will ionize the circumburst gas on short time
scales, thus decreasing the effective optical depth with time
\citep{pl98}. The detection of a transient iron edge in GRB990705
\citep{afv+00} and the decrease of the column density from the
prompt to the afterglow phases in GRB980329 \citep{fac+00} and
GRB010222 \citep{iz+01} are both consistent with this scenario.
%Curiously enough, these two
%bursts are not dark, arguing for a supplemental source of
%extinction in dark GRB's.
The variable size of the observable fireball that increases with
the inverse of the bulk Lorentz factor can also produce variations
of the column density, if the medium surrounding the source is not
homogeneous. In this regard, two X-ray afterglows show some,
admittedly marginal, evidence of variability of $N_H$
\citep{pcf+99,yym+01}.
 In conclusion, the future of the investigations
 of {\it dark} GRBs looks particularly bright.

\acknowledgments BeppoSAX is a program of the Italian space agency
(ASI) with participation of the Dutch space agency (NIVR). We
would like to thank H. Tananbaum and the Chandra team as well as
E. Costa, M. Feroci, J. Heise and the other members of the
BeppoSAX team for the support in performing the  observations with
these satellites and an anonymous referee for useful suggestions.
GG acknowledge support under NASA grant G00-1010X. MRG
acknowledges support under NASA contract NAG8-39073 to the Chandra
X-Ray Center. Part of the optical observations  are based on
observations made
 with the Danish 1.54-m telescope at ESO, La Silla, Chile.
 This research was supported by the Danish Natural Science Research
 Council through its Centre for Ground-based Observational Astronomy.
 Based in part on observations made with the Nordic Optical Telescope,
 operated  on the  island  of  La Palma  jointly  by Denmark,  Finland,
 Iceland, Norway, and Sweden, in  the Spanish Observatorio del Roque de
 los Muchachos of the Instituto de Astrof\'{\i}sica de Canarias.  Some of
 the data presented  here have been taken using  ALFOSC, which is owned
 by  the  Instituto de  Astrof\'{\i}sica  de  Andaluc\'{\i}a (IAA)  and
 operated at  the Nordic Optical Telescope under  agreement between IAA
 and    the    NBIfAFG   of    the    Astronomical   Observatory    of
 Copenhagen.  Partially  based  on  ESO  VLT  programme  66.A-0386(A),
 Cerro Paranal, Chile.

%\normalsize
%\bibliographystyle{apj}
%\bibliography{C:/LUIGI/PAPERS/Bibliography/journals_apj,C:/LUIGI/PAPERS/Bibliography/grbrefs,C:/LUIGI/PAPERS/Bibliography/000210,C:/LUIGI/PAPERS/Bibliography/jets}

% figure, plot the data

%\clearpage

%\begin{figure}
%  \plotone{radio.ps}
%\caption{The discovery image for the radio transient of \grb\
%  taken with the VLA on 2000 February 18.95 UT. The 50\arcsec\ error
%  radius of the X-ray afterglow detected by {\em BeppoSAX} is
%  indicated by the large error circle.  The 0.6\arcsec\ {\em Chandra}
%  error radius is indicated by the smaller circle. Radio contours are
%  plotted as $-$3.5, 3.5 and 4.5 times the rms noise of 21 $\mu$Jy
%  beam$^{-1}$.  The size and shape of the 5.8\arcsec\ $\times$
%  2.1\arcsec\ synthesized beam is shown in the left-hand corner.  {\bf
%    Dale. THIS WILL HAVE TO BE RE-DONE USING the final
%     {\em Chandra}  and radio positions}
%} \label{fig:radio}
%\end{figure}

%\begin{figure}
%\plotone{z1xi0and400mod.ps}
%\epsscale{0.9} \plotone{f8.ps} \caption{ Best fit models of the
%X-ray afterglow of GRB\,000210 assuming its redshift $z$=5, as
%would be observed at $z$=1. The dashed line is the case of a
%neutral absorber with $N_H=9\times10^{22}$ cm$^{-2}$; the
%continuous line is for an ionized absorber with $\xi=400$ and
%$N_H=16\times10^{22}$ cm$^{-2}$ } \label{fig:ionized}
%\end{figure}

\clearpage
\begin{deluxetable}{lcccllc}
\tabcolsep0in\footnotesize
%\tablewidth{\hsize}
\tablewidth{0in} \tablecaption{Optical Observations of
\grb\label{tab:Table-Optical}} \tablehead { \colhead {Epoch} &
\colhead {Filter}  & \colhead {Exposure} & \colhead {Seeing} &
\colhead {Telescope} &
\colhead {Magnitude} \\
\colhead {(UT)}      & \colhead { } & \colhead {Time (s)} &
\colhead {(arcsec)} & \colhead { } & \colhead { } } \startdata
Feb.  10.88 - 10.90  &  R  &  3x300      & 1.2 &  NOT  & $>23$\\
Feb.  11.03 - 11.08  &  R  &  10x300     & 1.6 & 1.54D & $23.5\pm0.2$\\
Feb.  14.02 - 14.03  &  R  &  600        & 1.9 & 1.54D & $>22.6$ \\
May    5.42 -  5.44  &  R  &  2x600      & 2.3 & 1.54D & $>22$ \\
Aug.  22.29 - 22.41  &  R  &  7x900      & 2.2 & 1.54D & $23.47\pm0.10$\\
Aug.  23.23 - 23.29  &  R  &  5x900      & 2.3 & 1.54D & "\\
Aug.  24.23 - 24.30  &  R  &  4x900      & 3.0 & 1.54D & "\\
Aug.  26.29 - 26.43  &  V  &  9x900      & 1.5 & 1.54D & $24.09\pm0.15$\\
Aug.  27.21 - 27.24  &  I  &  2x900      & 1.4 & 1.54D & $22.60\pm0.12$\\
Aug.  28.21 - 28.24  &  I  &  2x1200     & 1.4 & 1.54D & "\\
Aug.  29.21 - 29.30  &  I  &  7x1200     & 1.1 & 1.54D & "\\
Aug.  30.22 - 30.24  &  I  &  1200       & 1.1 & 1.54D & "\\
Aug.  31.21 - 31.24  &  B  &  2x1200     & 1.5 & 1.54D & $25.1\pm0.7$\\
Oct.  25.24 - 25.24  &  R  &  300     & 0.7 & VLT & $23.46\pm0.10$\\
\enddata
\tablecomments{R and I magnitudes from Aug. 22 to Aug. 24 and from
Aug.27 to Aug.30 have been derived by summing the images obtained
in those nights }
\end{deluxetable}

\newpage
\begin{deluxetable}{lccccccc}
\tabcolsep0.05in\footnotesize
%\tablewidth{\hsize}
\tablewidth{0in} \tablecaption{Positions and magnitudes of the
secondary standards reported in Fig.\ref{fig:optical}
\label{tab:secondary}} \tablehead{ \colhead {ID}      &
\colhead{$\alpha_{2000}$} & \colhead {$\delta_{2000}$} & \colhead
{I} &\colhead {R} & \colhead {V} & \colhead {B} & \colhead {U} }
\startdata
1& 01:59:10.056& -40:37:15.11& 17.21$\pm$  0.02& 17.71$\pm$ 0.02& 18.20$\pm$ 0.02& 19.01$\pm$0.03&19.64$\pm$0.08\\
2& 01:59:17.877& -40:37:20.58& 16.51$\pm$  0.02& 16.91$\pm$ 0.02& 17.26$\pm$ 0.02& 17.81$\pm$0.03&17.88$\pm$0.05\\
3& 01:59:16.762& -40:37:51.37& 15.93$\pm$  0.02& 17.00$\pm$ 0.02& 18.00$\pm$ 0.02& 19.40$\pm$0.03&20.71$\pm$0.12\\
4& 01:59:12.370& -40:37:57.60& 18.09$\pm$  0.03& 19.15$\pm$ 0.03& 20.18$\pm$ 0.03& 21.54$\pm$0.04&23.11$\pm$0.25\\
5& 01:59:04.478& -40:38:19.70& 17.34$\pm$  0.02& 17.64$\pm$ 0.02& 17.86$\pm$ 0.02& 18.16$\pm$0.03&17.99$\pm$0.06\\
6& 01:59:06.665& -40:38:21.51& 16.07$\pm$  0.02& 16.52$\pm$ 0.02& 16.89$\pm$ 0.02& 17.47$\pm$0.03&17.49$\pm$0.04\\
7& 01:59:16.673& -40:40:20.09& 16.60$\pm$  0.02& 16.74$\pm$ 0.02& 16.78$\pm$ 0.02& 16.83$\pm$0.02&17.07$\pm$0.03\\
\enddata
\end{deluxetable}

\newpage
\begin{deluxetable}{lccccr}
\tabcolsep0in\footnotesize
%\tablewidth{\hsize}
\tablewidth{0in} \tablecaption{Radio Observations of
\grb\label{tab:Table-Radio}} \tablehead { \colhead {Epoch}      &
\colhead {$\Delta$t}  & \colhead {$\nu$} & \colhead {Telescope} &
\colhead {~Array} &
\colhead {S$\pm\sigma$} \\
\colhead {(UT)}      & \colhead {(days)} & \colhead {(GHz)} &
\colhead {} & \colhead {} & \colhead {($\mu$Jy)} } \startdata 2000
Feb. 10.98    &   0.62 & 8.46  & VLA  & B   &   34$\pm$70 \nl 2000
Feb. 12.32    &   1.96 & 8.70  & ATCA & 6A  &   $-4\pm$59 \nl 2000
Feb. 14.90    &   4.54 & 8.46  & VLA  & B   &   15$\pm$37 \nl 2000
Feb. 15.03    &   4.67 & 8.46  & VLA  & CnB & $-34\pm$42 \nl 2000
Feb. 18.95    &   8.59 & 8.46  & VLA  & CnB &  93$\pm$21 \nl 2000
Feb. 26.95    &  16.59 & 8.46  & VLA  & CnB &   20$\pm$16 \nl 2000
Mar. 03.92    &  21.56 & 8.46  & VLA  & CnB &  58$\pm$33 \nl 2000
Mar. 27.81    &  45.45 & 8.46  & VLA  & C   &  45$\pm$45 \nl 2000
May  28.73    & 107.37 & 8.46  & VLA  & C   &  48$\pm$26 \nl 2000
Jun. 24.69    & 134.33 & 8.46  & VLA  & D   &  $-1\pm$34  \nl
\enddata \tablecomments{The columns are (left to right), (1) UT date
  for each observation, (2) time elapsed since the $\gamma$-ray burst,
  (3) observing frequency, (4) telescope name, (5) the array
  configuration, and (6) peak flux density at the best fit position of
  the radio transient, with the error given as the root mean square
  noise in the image.}
\end{deluxetable}

%\begin{table}
%\caption[] {Optical photometry}
%\begin{flushleft}
%\begin{tabular}{lllll}
%\noalign {\hrule} \noalign {\medskip}
%Date    &       Filter & Magnitude  &       Telescope \\
%Feb.10.88-10.9  &   R  &     $>23?$      &   NOT \\
%Feb.11.03-11.08 &   R   &   $23.5\pm0.2$&   1.54D \\
%Feb.11??-??     &   R   &   $23.46\pm0.16$& 2.2m \\
%Feb.11??-??     &   V   &   $24.3\pm0.2 $&  2.2m \\
%Feb.11??-??     &   B   &   $24.2\pm0.2$&   2.2m \\
%Feb.11??-??     &   I   &   $>22.5$&        2.2m \\
%Feb.12??-??     &   R   &   $23.5\pm0.25$&  NTT \\
%Feb.12??-??     &   V   &   $24.2\pm0.2$ &  NTT \\
%Mar.3.??-??     &   R   &   $23.53\pm0.15$& 3.6m \\
%Aug.22-24       &   R   &   $23.47\pm0.10$&  1.54D \\
%Aug.26          &   V   &   $24.09\pm0.05$& 1.54D \\
%Aug.27-30       &   I   &   $22.70\pm0.06$& 1.54D \\
%Aug.31          &   B   &   $>24.5$         & 1.54D \\
%\noalign {\hrule} \noalign{\medskip} \noalign{\noindent Note: }
%\end{tabular}
%\end{flushleft}
%\end{table}

%\clearpage
%\begin{center}
%\begin{figure}[H]
% \caption{\label{fig:host} The figure shows a blow up of the co-added R-band
%   image taken on Aug 22.29-- 24.30 UT.  The plot shows the position of the
%   improved Chandra X-ray circle \cite{Garmire00} and the optical candidate
%   \cite{Gorosabel00b}.}
% \resizebox{\hsize}{!}{\includegraphics[angle=90]{fig2.ps}}
%\end{figure}
%\end{center}

\end{document}